\documentclass[prd,aps,preprint,tightenlines,superscriptaddress,showpacs]{revtex4}
\usepackage{epsfig}
\usepackage{amsmath}

\newcommand{\beq}[1]{\begin{equation} \label{#1}}
\newcommand{\eeq}{\end{equation}}
\newcommand{\beqar}[1]{\begin{eqnarray} \label{#1}}
\newcommand{\eeqar}{\end{eqnarray}}
\newcommand{\beqs}{\begin{eqnarray}}
\newcommand{\eeqs}{\end{eqnarray}}
\newcommand{\bra}[1]{\langle {#1}|}
\newcommand{\ket}[1]{|{#1}\rangle}

\newcommand{\nid}{\noindent}
\newcommand{\non}{\nonumber}
\newcommand{\bib}{\bibitem}

\newcommand{\etc}{{\it et al }}

\begin{document}

\title{General Parton Distributions and \\
Counting of Helicity-Flip Nucleon Form Factors
\footnote{Talk given at OCPA04, Joint International Conference of
Chinese Physicists Worldwide, Shanghai, PRC.}}
\author{Zhang Chen}
\email{chenz@mville.edu}
\affiliation{Department of Physics, Manhattanville College, Purchase, NY 10577, USA}
\date{\today}

\vspace{0.5in}

\begin{abstract}
We give a brief overview of the General Parton Distributions(GPDs), including their
properties, connections to conventional quantities, and relationships to physical
observables. We also perform a systematic analysis on the non-forward matrix elements
of twist-two quark and gluon helicity-flip operators and explain their relationships
to the GPDs. We systematically count the number of independent nucleon form factors
in non-forward scattering (of matrix elements of these operators), by matching the allowed
quantum numbers with their crossing channel counterparts and time reversal symmetry
considerations \cite{ref:zc-count} (a method developed in \cite{ref:count}).
We finally analyze and write down the form factor expansion of the quark operator
matrix elements.
%in terms of tensorial/Lorentz structure, kinematic factors, and the independent form factors.
\end{abstract}


% 11.80.Cr: Kinematical properties (helicity and invariant amplitudes,
% kinematic singularities, etc.)
% 11.30.Er: Charge conjugation, parity, time reversal, and other
% discrete symmetries
% 11.40.-q: Currents and their properties
% 14.20.Dh: Properties of protons and neutrons
% 13.60.-r Photon and charged-lepton interactions with hadrons

\maketitle

\section{Introduction}

One of the most important frontiers in strong interaction physics is
the study of the structure of the nucleon. Many unanswered questions
still exist due to the non-perturbative nature of the
bound state problem in Quantum ChromoDynamics (QCD). However, recent
theoretical and experimental efforts on the so-called General Parton
Distributions (GPDs) have shed new light on the problem, especially
on the possibilities of a three-dimensional view and a complete description
of the nucleon (eg, see \cite{ref:ofpd, ref:Diehl, ref:zc-spd} and
many others).

The GPDs generalize and interpolate between the ordinary Parton Distribution
Functions (PDFs) and elastic form factors--both having been
studied extensively for many years--and therefore contain rich structural
information. They can be accessed in high-energy (exclusive) diffractive
processes in which the nucleon recoils elastically after receiving a
non-zero momentum transfer in the so-called deeply virtual limit, eg,
Deeply Virtual Compton Scattering (DVCS) and diffractive electro-production
of vector mesons. From a more theoretical point of view, GPDs are closely
related to the matrix elements of quark and gluon operators in QCD, through
generalized nucleon form factors, the knowledge of which is essential
to fully describe the nucleon. In particular, the matrix elements of those
operators of twist-two are of the most importance and interest,
because they usually give the leading contribution (hence the synonym
leading-twist) in appropriate hard processes, have clear physical
interpretation (eg, corresponding to the energy-momentum tensor), and
are often more accessible to experimental measurement and relatively simple.

The enumeration of independent nucleon form factors of twist-two operators
and the expansion into form factors of non-forward matrix elements of these
operators are among the essential understandings of the problem. We
extend such studies, following a method based on the partial wave formalism
and crossing symmetry (first developed in \cite{ref:count}), to helicity-flip
twist-two quark and gluon operators \cite{ref:flip}, by systematically enumerate
the number of independent hadronic form factors of both and write down the
form factor expansion of the general quark operators \cite{ref:zc-count}.

\section{General Parton Distributions} \label{sec:GPDs}

Generalized Parton Distributions represent the low-energy (soft) internal
structure of the nucleon, in three dimensions. They connect and are
generalizations of Feynman PDFs and elastic electromagnetic form factors.
The PDFs usually result from the overlap of hadron wave functions and
contain information on the longitudinal momenta and polarizations carried
by various partons in a fast moving hadron. On the other hand, the
traditional form factors contain information on the transverse momenta
of partons, often in the form of sum rules related to, for example,
charges, local currents and the energy-momentum tensor (of QCD).
The GPDs, on the other hand, in general have their physical interpretations
given in terms of probability amplitudes and/or parton correlation
functions, and have simple physical significance in light-cone coordinates
(or the infinite momentum frame).

\subsection{Forward and Non-forward Scenarios}

In the forward scenario, eg, in deep inelastic
scattering (DIS), one relates the imaginary part of a forward
(Compton) scattering amplitude to the cross section via the
optical theorem. The factorization of the forward amplitude
into the hard scattering part and the soft physics is achieved
through the formal approach of operator product expansion (OPE)
\cite{ref:Collins}.
The hard scattering is calculated order by order in perturbation
theory and the soft part is parameterized as the PDFs, which
are essentially matrix elements of light-cone bi-local quark
and gluon operators between equal momentum states. A renormalization
group equation governs the factorization scale dependence of
the PDFs/matrix elements, while the cross section is eventually
related to (usually a linear combination of) the PDFs.

\begin{figure}[h]
\centerline{\epsfig{file=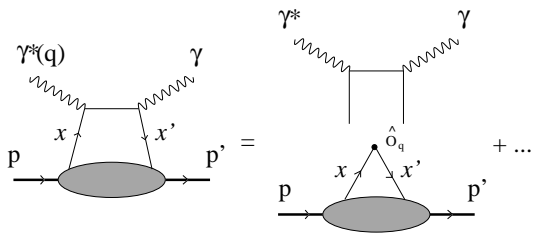,width=7cm}}
\vspace{4pt}
\caption{Non-forward OPE (with only (lowest) $q-q$ diagram shown)}
\label{fig:ope}
\end{figure}

The GPDs arise mostly in non-forward scattering processes, and
among such processes DVCS is the cleanest of all. Various studies
\cite{ref:nfvcs} have shown that even in such cases, with both
longitudinal and transverse momentum transfer, factorization is
still valid and non-forward OPE can be similarly performed.
That is, the non-forward amplitude can be
factorized into the hard scattering part, calculable order by
order in perturbation theory, and the soft part, now parameterized
as GPDs (with proper insertions of operators), as shown in figure
\ref{fig:ope}.

\nid In particular, the lower parts of the above graph give rise
to four general quark distributions (from two different operators):
$\Big(H_{q}(x, \xi, t), \; E_{q}(x, \xi, t) \Big)$ (quark spin sum)
and $\Big(\tilde H_{q}(x,\xi, t), \; \tilde E_{q}(x, \xi, t)\Big)$
(spin difference), where, the now standard kinematical variables
used in discussions of GPDs (through for example DVCS) are defined as
(see, eg, \cite{ref:ofpd, ref:flip})
\begin{eqnarray}
\overline{P}^\mu &=& (P'+P)^\mu/2 \,=\, p^\mu + (\bar M^2/2) n^\mu\ , \nonumber \\
\Delta^\mu &=& P'^\mu-P^\mu \,=\,-2\xi(p^\mu-(\bar M^2/2) n^\mu)
+ \Delta_\perp^\mu\ ,\nonumber \\
\bar M^2 &=& M^2 -\Delta^2/4 \ , \label{eqn:kinematics}
\end{eqnarray}

\nid and $x$ and $\xi$ are longitudinal momentum fractions referring
to $\overline{P}$ (The initial nucleon and parton have longitudinal
momentum fractions $1+\xi$ and $x+\xi$, respectively.).

\subsection{Connection to Other Observables}

GPDs display characteristics of both the forward parton distributions
and nucleon form factors. They reduce to the DIS structure functions in
the forward limit of $ ( \; P = P', ~~~ \xi = t = 0 \;)$
$$
\hspace*{-2cm} {\rm eg,} ~~~~~ H_q(x,0,0) = q(x),~~~ \tilde{H}_q(x,0,0) = \Delta q(x)
$$

\nid where $q(x)$ and $\Delta q(x)$ are forward quark (spin independent)
 and quark helicity (spin dependent) distributions.

GDPs are also closely related to the elastic form factors. Their
first moments (integration over $x$ of the GPDs) are related
to the nucleon form factors by the sum rules:
\begin{eqnarray}
\int^1_{-1} dx H_q(x,\xi,t) = F_1(t) \ , &\;\;&
\int^1_{-1} dx E_q(x,\xi,t) = F_2(t) \ ,     \nonumber     \\
&& \nonumber \\
\int^1_{-1} dx \tilde{H}_q(x,\xi,t) = G_A(t)\ , &\;\;&
\int^1_{-1} dx \tilde{E}_q(x,\xi,t) = G_P(t)\   \nonumber.
\end{eqnarray}

\nid where $F_1$ and $F_2$ are quark Dirac and Pauli form factors (for
electromagnetic current), respectively, and $G_A$ and $G_P$ the quark
axial and pseudoscalar form factors (for axial current), respectively.
Higher moments of GPDs are related to (usually linear combinations of)
general nucleon form factors of general parton (local) operators \cite{ref:ofpd,ref:Diehl}.

Most interestingly, the GPDs contain rich information on the spin structure
of the nucleon, in particular, about the orbital motion of partons in
a (polarized) nucleon. Nucleon spin flip needs orbital angular momentum,
and GPDs (through DVCS) is the only place so far probing it--through
Ji's Sum Rule \cite{ref:spin}
$$
{1 \over 2} \int^1_{-1} \; dx \; x \; (H_q + E_q) = J_q(t)
$$

\nid where $J_q(0)$  is the {\it Total Angular Momentum} of quarks
(spin + orbital). A similar result holds for the gluons and one finds
$$
J_q(0) + J_g(0) = {1 \over 2} .
$$

\nid Indeed, it is believed that GPDs offer an opportunity of a more
complete description of the nucleon wave-function. %\cite{ref:JCPeng}

\subsection{General Parton Operators and GPDs}

In the forward case, the irreducible matrix elements
of the parton operators, obtained after factorizing out the
tensorial/Lorentz structure and kinematic factors, are combinations
of moments of the conventional PDFs, and can be used to define
the PDFs (\cite{ref:Collins}). In the non-forward case,
similarly, matrix elements of (twist-two) parton operators are expanded
into general nucleon form factors (and kinematic/Lorentz factors).
Moments of GPDs are linear combinations of these form factors,
that is, they are polynomials in the general nucleon form factors
in powers of $\xi$. These general "sum rules" can be used to
define the GPDs. From the point view of the low-energy nucleon
structure, the GPDs can be considered as the generating functions
for the form factors of the twist-two operators. The following is the
list of twist-two quark and gluon operators and their corresponding
GPDs \cite{ref:ofpd, ref:Diehl, ref:zc-count}:
%\newpage
\begin{tabbing}
Social Se \= So \= Social Security aaaaaaaaaaa
\= Social\= Social Security \= \kill
${\cal O}^{\mu_1\cdots\mu_n}_q$
    \> $=$ \>
    $\overline{\psi}_q i\stackrel{\leftrightarrow}{\cal D}^{(\mu_1}
        \cdots  i\stackrel{\leftrightarrow}{\cal D}^{\mu_{n-1}}
        \gamma^{\mu_n)} \psi_q $
    \> $\longleftrightarrow$ \>
    $\Big(H_{q}(x, \xi, t), \; E_{q}(x, \xi, t) \Big)$
    \\
%\>\\
$\tilde {\cal O}^{\mu_1\cdots\mu_n}_q$
    \> $=$ \>
    $\overline \psi_q i\stackrel{\leftrightarrow}{\cal D}^{(\mu_1}
        \cdots  i\stackrel{\leftrightarrow}{\cal D}^{\mu_{n-1}}
        \gamma^{\mu_n)} \gamma_5 \psi_q $
    \> $\longleftrightarrow$ \>
    $\Big(\tilde H_{q}(x,\xi, t), \; \tilde E_{q}(x, \xi, t)\Big)$
    \\
%\>\\
${\cal O}^{\mu_1\cdots\mu_n}_g$
    \> $=$ \>
    $F^{(\mu_1\alpha} i\stackrel{\leftrightarrow}{\cal D}^{\mu_2}
        \cdots  i\stackrel{\leftrightarrow}{\cal D}^{\mu_{n-1}}
        F_\alpha^{~\mu_n)}$
    \> $\longleftrightarrow$ \>
    $\Big( H_{g}(x,\xi, t), \; E_{g}(x, \xi, t)\Big)$
    \\
%\>\\
$\tilde {\cal O}^{\mu_1\cdots\mu_n}_g$
    \> $=$ \>
    $F^{(\mu_1\alpha} i\stackrel{\leftrightarrow}{\cal D}^{\mu_2}
        \cdots  i\stackrel{\leftrightarrow}{\cal D}^{\mu_{n-1}}
        i\tilde F_\alpha^{~\mu_n)}$
    \> $\longleftrightarrow$ \>
    $\Big(\tilde H_{g}(x,\xi, t), \; \tilde E_{g}(x, \xi, t)\Big)$
    \\
%\>\\
${\cal O}^{\mu_1\cdots\mu_n\alpha}_{qT}$
    \> $=$ \>
    $\overline \psi_q i\stackrel{\leftrightarrow}{\cal D}^{(\mu_1}
        \cdots  i\stackrel{\leftrightarrow}{\cal D}^{\mu_{n-1}}
        \sigma^{\mu_n)\alpha} \psi_q $
    \> $\longleftrightarrow$ \>
    $\Big(H_{Tq}(x,\xi, t), \; E_{Tq}(x, \xi, t), \; \tilde H_{Tq}(x,\xi, t),
        \; \tilde E_{Tq}(x, \xi, t) \Big)$
    \\
%\>\\
${\cal O}^{\mu_1\cdots\mu_n\alpha\beta}_{gT}$
    \> $=$ \>
    $F^{(\mu_1\alpha} i\stackrel{\leftrightarrow}{\cal D}^{\mu_2}
        \cdots  i\stackrel{\leftrightarrow}{\cal D}^{\mu_{n-1}}
        F^{\mu_n)\beta}$
    \> $\longleftrightarrow$ \>
    $\Big(H_{Tg}(x,\xi, t), \; E_{Tg}(x, \xi, t), \; \tilde H_{Tg}(x,\xi, t),
        \; \tilde E_{Tg}(x, \xi, t) \Big)$
    \\
\end{tabbing}

\nid We concentrate on the last two--each of which corresponds
to $4$ (helicity-flip) GPDs. For the lowest spin, the helicity-flip
GPDs are defined explicitly by \cite{ref:flip,ref:Diehl,ref:Diehl_Ori}
\begin{eqnarray}
    \int  {d\lambda \over 2\pi} e^{i\lambda x}
      \langle P'S'|\bar\psi_q(-\frac{1}{2}\lambda n)\sigma^{\mu \nu}
            \psi_q(\frac{1}{2}\lambda n)|PS \rangle
      &=& H_{Tq}(x,\xi) \,
        \bar U(P'S')\sigma^{\mu \nu} \, U(PS) \non \\
    + \; \tilde H_{Tq}(x,\xi) \, \bar U(P'S')
        {\overline{P}^{[\mu} i\Delta^{\alpha ]} \over M^2} \,U(PS)
    & + & E_{Tq}(x,\xi) \, \bar U(P'S')
          {\gamma^{[\mu}i\Delta^{\nu]} \over M} \, U(PS) \non \\
    + \; \tilde E_{Tq}(x,\xi) \!\!\!\!&&\!\!\!\! \bar U(P'S')
          {\gamma^{[\mu}i\overline{P}^{\nu]} \over M} \, U(PS) + ...\ , \non \\
    {1\over x}\int { d\lambda\over 2\pi} e^{i\lambda x}
        \langle P'S'|F^{(\mu\alpha}(-{\lambda\over2}n)
       F^{\nu\beta)}({\lambda\over2}n)|PS\rangle
        &=& H_{Tg}(x,\xi) \, \bar U(P'S'){\overline{P}^{([\mu} i\Delta^{\alpha ]}
            \sigma^{\nu\beta)} \over M} \,U(PS) \non \\
    + \; \tilde H_{Tg}(x,\xi) \, \bar U(P'S')
        {\overline{P}^{([\mu} \Delta^{\alpha ]}\over M}
        {\overline{P}^{[\nu}\Delta^{\beta])} \over M^2} \, U(PS)
    & + & E_{Tg}(x,\xi) \, \bar U(P'S')
        {\overline{P}^{([\mu} \Delta^{\alpha ]}\over M}
        {\gamma^{[\nu}\Delta^{\beta])} \over M} \, U(PS) \non \\
    + \; \tilde E_{Tg}(x,\xi) \!\!\!\!&&\!\!\!\! \bar U(P'S')
        {\overline{P}^{([\mu} \Delta^{\alpha ]}\over M}
        {\gamma^{[\nu}\overline{P}^{\beta])} \over M} \, U(PS) + ... \,
\label{eqn:lowest-op}
\end{eqnarray}

\nid where each distribution is also implicitly dependent on
$Q^2$ and $t=\Delta^2$. In the first equation, $[\mu\nu]$ denotes
anti-symmetrization of the two indices and the ellipses represent
higher twist structures. The gauge link between the quark fields is
not explicitly shown. In the second equation $[\mu\alpha]$ and
$[\nu\beta]$ are antisymmetric pairs and $(\cdots)$ signifies
symmetrization of the two and removal of the trace.

\section{Counting of Twist-2 Helicity-Flip Nucleon Form Factors}

\subsection{The General Counting Method}

A general method was developed by Ji and Lebed \cite{ref:count}
to count the number of independent nucleon form factors using
overall CPT invariance and crossing symmetry. It is a basic property
of relativistic quantum field theory that the number of independent
amplitudes is the same in all crossed channels. Therefore, in the
direct channel one uses parity and time reversal invariance for
non-forward matrix elements of the parton operators
$\bra{\bar{P}} {\cal O} \ket{P}$, while in the crossed channel
one uses parity and charge conjugation invariance for
creation of a particle-antiparticle pair
$\bra{P\bar{P}} {\cal O} \ket{0}$. The two resulting types
of constraints are equivalent due to overall CPT invariance,
leading to same structure for allowed form factor decompositions.
Thus by matching the allowed quantum numbers of the two
channels one can systematically count the number of independent
nucleon form factors. We apply this method to the helicity-flip
operators/nucleon form factors (in eqn. \ref{eqn:lowest-op}).

\subsection{Counting of Independent Nucleon Form Factors}

The quark helicity-flip operator ${\cal O}^{\mu_1\cdots\mu_n\alpha}_{qT}$
transforms as representations of Lorentz group
($\frac{n-1}{2}$, $\frac{n+1}{2}$) and (plus) ($\frac{n+1}{2}$, $\frac{n-1}{2}$).
As a tensor of $(n,1)$ (with $n$ symmetric and $1$ pair of anti-symmetric
indices), the generic tensorial counting gives its number of
independent elements as $ 2 \times n(n+2)$ (for details on
notation and tensorial counting, see \cite{ref:Hamermesh, ref:Landau, ref:zc-count}).

For the matrix elements in the crossed channel,
$\bra{P\bar{P}} {\cal O}^{\mu_1 ... \mu_n \nu} \ket{0}$,
the particle and antiparticle pair's $J^{PC}(L)$ values are
$J = L+S$, $P=(-1)^{L+1}$, $C=(-1)^{L+S}$, $S=0, 1$. In terms
of $J=n$, we have $S=0$: $L=J$, $P=(-1)^{J+1}$, $C=(-1)^J, (-1)^{J+1}$,
while $S=1$: $L=J\!\pm\!1$, $P=(-1)^J$, $C=(-1)^J$.
On the other hand, for the matrix elements in the direct channel,
$\bra{\bar{P}} {\cal O}^{\mu_1 ... \mu_n \nu} \ket{P}$, the representation
$(A, B)$ and $(B,A)$ have angular momentum $J = |A-B|, |A-B|+1, ..., A+B$
(since $\vec{J} = \vec{A} + \vec{B}$). Since parity transforms
$A \leftrightarrow B$, for each $J$, both $J^\pm$ are allowed.
The charge conjugation $C$ of $\gamma^\mu$, (each) $i
\stackrel{\leftrightarrow}{\cal D}^\mu$, and $\sigma^{\mu\nu}$ are
all $-1$. Thus we have $J = 1, 2, ...\;, n$, $P=\pm$, $C = (-1)^n$
(with $L = J \pm 1$).

From the above analysis, we obtain the results in Table
\ref{tab:quark}, including lists of allowed quantum numbers in
both channels, the allowed quantum number structure from matching
between the two channels, and the enumeration of independent form
factors for a rank $r=n+1$ twist-two quark helicity-flip operator.

\begin{table}[ht]
%\vspace{4pt}
\caption{The enumeration of independent form factors of helicity-flip
quark operators. In the table, $J_{max}=r-1\equiv n$ and the $(\times 2)$
represents the two different $L$ ($=J\pm1$) values for each $J$.}
\label{tab:quark}
%
%\begin{center}
%
\vspace{12pt}
\begin{tabular}{||c|c|c|c|c||}
\hline\hline
&&&&\\
$J_{max}$
& $ P\bar{P} \; (J^{PC}(L))$ & ${\cal O}^{\mu_1...\mu_n \nu} \; (J^{PC})$
& $Matched \; J^{PC}$ & $Enumeration$\\
&&&&\\
\hline
$0$
& $~0^{++}(1), ~0^{-+}(0)$ & $~N/A$
& $N/A$ & $N/A$ \\
\cline{1-1} \cline{3-5}
$1$ & $~1^{++}(1), ~1^{+-}(1), $ & $~1^{+-}, ~1^{--}$
    & $~1^{+-}, ~1^{--} (\times 2) $ & $ (1+2) = 3 $ \\
    & $~1^{--}(0), ~1^{--}(2)$ & & & \\
%\hline
\cline{1-1} \cline{3-5}
$2$ & $~2^{++}(1), ~2^{++}(3),$ & $~1^{++}, 1^{-+}, 2^{++}, 2^{-+}$
    & $~1^{++}; ~2^{++} (\times 2), ~2^{-+}$ & $~(1) + (1+2) = 4$ \\
    & $~2^{-+}(2), ~2^{--}(2)$ & & & \\
%\hline
\cline{1-1}\cline{3-5}
$3$ & $~3^{++}(3), ~3^{+-}(3),~$ & $~1^{+-}, 1^{--}, 2^{+-}, 2^{--},$
    & $~1^{+-}, ~1^{--} (\times 2); ~2^{--};$ & $~(1+2) + (1) + (1+2) = 7 $ \\
    & $~3^{--}(2), ~3^{--}(4)$ & $~3^{+-}, ~3^{--}$
    & $~3^{+-}, ~3^{--} (\times 2)$ & \\
%\hline
\cline{1-1} \cline{3-5}
$4$ & $~4^{++}(3), ~4^{++}(5),$ & $~1^{++}, 1^{-+}, 2^{++}, 2^{-+},$
    & $~1^{++}; ~2^{++} (\times 2), ~2^{-+};$ & $~(1) + (1+2) + (1)$ \\
    & $~4^{-+}(4), ~4^{--}(4)$ & $~3^{++}, 3^{-+}, 4^{++}, 4^{-+}$
    & $~3^{++}; ~4^{++} (\times 2), ~4^{-+}$ & $+ (1+2) = 8$ \\
%\hline
\cline{1-1} \cline{3-5}
$\cdots$ & $\cdots$ & $\cdots$ & $\cdots$ & $\cdots$ \\
%\hline
\cline{1-1} \cline{3-5}
&&&&\\
$odd$
    & $n^{--}(n\!\pm\!1),$
    & $~1^{+-}, ~1^{--}, ~... \;,$
    & $~1^{+-}, ~1^{--} (\times 2); ~2^{--}; $
    & $ ~(1\!+\!2)+(1)\!+\! ... \!+\!(1\!+\!2) = $ \\
$n$ & $n^{++}(n), ~n^{+-}(n)$
    & $~n^{+-}, ~n^{--}$ & $~...\; ~n^{+-}, ~n^{--} (\times 2)$
    & $~3 \frac{n+1}{2} \!+\! \frac{n-1}{2} \!=\! 2n\!+\!1 (=\! 2r\!-\!1)$ \\
%\hline
&&&&\\
\cline{1-1} \cline{3-5}
&&&&\\
$even$
    & $n^{++}(n\!\pm\!1),$
    & $~1^{++}, ~1^{-+}, ~... \;,$
    & $~1^{++}; ~2^{++} (\times 2), ~2^{-+}; $
    & $ ~(1) \!+\! (1\!+\!2) \!+\! ... \!+\!(1\!+\!2) \!=\!$ \\
$n$ & $~n^{-+}(n), ~n^{--}(n)$
    & $~n^{++}, ~n^{-+}$ & $~...\;; ~n^{++} (\times 2), ~n^{-+}$
    & $\frac{n}{2} \!\times\! 3 + \frac{n}{2} = 2n \, (= 2r \!-\! 2) $ \\
&&&&\\
\hline\hline
\end{tabular}
%\end{center}
\end{table}

In the case of the gluon helicity-flip operators
$\; {\cal O}^{\mu \alpha \nu \beta \mu_1 ... \mu_n}_{gT}$ ( = $F^{\mu \alpha}
        i \stackrel {\leftrightarrow}{\cal D}^{\mu_1} \cdots \;
        i \stackrel{\leftrightarrow}{\cal D}^{\mu_n} F^{\nu \beta}$ ),
they are $(n+2, 2)$ tensors and transform under representations
$(\frac{n}{2}, \frac{n+4}{2})$  and $(\frac{n+4}{2}, \frac{n}{2})$,
with angular momentum values $J = 2, 3, \cdots, n\!+\!2$. At the same
time, both parity values are allowed, and the bilinear gluon fields
$FF$ have positive charge conjugation. Following similar steps as
in the quark case, the enumeration in Table \ref{tab:gluon}
is obtained ($r=n+4$ is the rank of the tensor operator).

\begin{table}[ht]
\vspace{6pt}
\caption{The enumeration of independent form factors of helicity-flip
gluon operators.}
\label{tab:gluon}
\begin{center}
\begin{tabular}{||c|c|c|c||}
\hline\hline
&&&\\
$n$ & ${\cal O}^{\mu
\alpha \nu \beta \mu_1 ... \mu_n}$
    & $ Matched \; (J^{PC}(L))$ & $Enumeration$  \\
&&&\\
\hline
$0$ & $~2^{++}, ~2^{-+}$
    & $~2^{++}(1), ~2^{++}(3), ~2^{-+}(2)$ & $3$\\
\hline
$1$ & $~2^{+-}, ~2^{--}, ~3^{+-}, ~3^{--}$
    & $~2^{--}(2), ~3^{+-}(3), ~3^{--}(2), ~3^{--}(4)$ & $1+3 =4$\\
\hline
$2$ & $~2^{++}, ~2^{-+}, ~3^{++}, ~3^{-+}$
    & $~2^{++}(1), ~2^{++}(3), ~2^{-+}(2), ~3^{++}(3)$ & $3+1+3$\\
        & $4^{++}, ~4^{-+}$
    & $~4^{++}(3), ~4^{++}(5), ~4^{-+}(4)$ & $\;\;\;\; = 7$\\
\hline
$3$ & $~2^{+-}, ~2^{--}, ~3^{+-}, ~3^{--}$
    & $~2^{--}(2), ~3^{+-}(3), ~3^{--}(2), ~3^{--}(4)$ & $1+3+1$ \\
        & $~4^{+-}, ~4^{--}, ~5^{+-}, ~5^{--}$
    & $~4^{--}(4), ~5^{+-}(5), ~5^{--}(4), ~5^{--}(6)$ & $\; +3 = 8$ \\
\hline
$\cdots$ & $\cdots$ & $\cdots$ & $\cdots$  \\
\hline
&&&\\
$odd \; n$ & $2^{+-}, ~2^{--}, ~... \;,$ & $ 2^{--}(2),
~3^{+-}(3), ~3^{--}\!\times\!2, \cdots,$
        & $(1+3)\times \frac{n\!+\!1}{2} $\\
    & $[n\!+\!2]^{+-}, ~[n\!+\!2]^{--}$
        & $[n\!+\!2]^{+-}(n\!+\!2), ~[n\!+\!2]^{--} \times 2$
        & $~=2(n\!+\!1)= 2r\!-\!6$\\
&&&\\
\hline
&&&\\
$even \; n$ & $2^{++}, ~2^{-+}, ~... \;,$ & $2^{++} \times 2,
~2^{-+}(2), \cdots,$
        & $ 3+(1+3)\!\times\!\frac{n}{2}$\\
    & $[n\!+\!2]^{++}, ~[n\!+\!2]^{-+}$ & $[n\!+\!2]^{++} \times 2, ~[n\!+\!2]^{-+}(n\!+\!2)$
        & $ ~= 2n\!+\!3 = 2r\!-\!5$ .\\
&&&\\
\hline\hline
\end{tabular}
\end{center}
\end{table}

\subsection{Form Factor Decomposition of Quark Operators}

Before we can write down the form factor decomposition of the quark operators,
we need to discuss whether time-reversal invariance (Hermiticity requirement)
would impose further constraints on the form factors.

The higher rank quark operators will have factors of $\overline{P}$ and $\Delta$
after taking matrix elements between $P'$ and $P$, coming from the covariant
derivatives \cite{ref:ofpd, ref:CZ}. From Hermiticity requirements, each factor
of $\Delta^\mu$ will introduce a factor of $-1$ while $P^\mu$ factor will not.
The overall sign factor from the covariant derivatives in the matrix element,
will be $(-1)^l$ with $l$ the number of factors of $\Delta$. On the other hand,
the Lorentz structures have the following behavior
$$\overline{U} [\gamma^\alpha, \gamma^{\mu_1}] U \sim \sigma^{\mu\nu}; ~~~
\overline{U} [\gamma^\alpha, \overline{P}^{\mu_1}] U \sim \gamma^\mu; ~~~
\overline{U} [\gamma^\alpha, \Delta^{\mu_1}] U \sim \gamma^\mu; ~~~~
\overline{U} [\overline{P}^\alpha, \Delta^{\mu_1}] U \sim \overline{U}U$$

\nid where the $\sim$ sign means having the same properties under time reversal
(hermitian, complex conjugate) and the first type of terms is odd ("-"), while
the rest three are even ("+"). These four are the only ones we need to
consider in this case (\cite{ref:ofpd, ref:Diehl, ref:zc-count}.
The operator ${\cal O}^{\alpha \mu_1 \mu_2 \cdots \mu_n}$ is odd ("-") overall.
Therefore to give the matrix elements the proper signs under time
reversal/Hermitian, namely overall odd ("-"), we have the
enumeration in Table \ref{tab:hermiticity} from these potentially additional
constraints for the
matrix element $\bra{P'} {\cal O}^{\alpha \mu_1 \mu_2 \cdots \mu_n} \ket{P}$.

\begin{table}[ht]
\vspace{6pt}
\caption{Enumeration from time-reversal/Hermiticity considerations}
\label{tab:hermiticity}
\vspace{12pt}
\begin{center}
\begin{tabular}{||c|c|c|c|c|c||}
\noalign{\vspace{-8pt}}
    Term & $~\hat{\mathcal{T}}~$
    & \multicolumn{4}{|c||}{Sign allowed from Factors of $\overline{P}$ and $\Delta$} \\
    \cline{3-6}
& & $~n\!=\!1~$ & $\cdots$ & $ \;\;\; n=2k+1$ & $ \;\;\; n=2k$ \\
\hline
$\overline{U} [\gamma^\alpha, \gamma^{\mu_1}] U$ & $ - $ & $
\;\; + $
    && $  \;\; (-1)^l,$ with & $ \;\; (-1)^l,$ with \\
&&&& $~l=0, 2, \cdots, 2k (= n\!-\!1)$ & $~l0, 2, \cdots, 2k\!-\!2 (= n\!-\!2)$ \\
(\# allowed) &&$\;\; (1)$&& $  \;\;\; (k+1 = \frac{n+1}{2})$ &$ \;\; (k = \frac{n}{2})$ \\
\hline
%&&&&&\\
%\hline
$\overline{U} [\gamma^\alpha, \overline{P}^{\mu_1}] U$ & $ + $ &
$ \;\; + $
    && $  \;\; (-1)^l,$ with & $ \;\; (-1)^l,$ with\\
&&&& $~l = 1, 3, \cdots, 2k\!-\!1 (= n\!-\!2)$ & $~l = 1, 3, \cdots, 2k\!-\!1 (= n\!-\!1)$ \\
(\# allowed) &&$\;\; (0)$&& $  \;\;\; (k = \frac{n-1}{2})$ &$ \;\; (k = \frac{n}{2})$ \\
\hline
%&&&&&\\
%\hline
$\overline{U} [\gamma^\alpha, \Delta^{\mu_1}] U$ & $ + $ & $
\;\; - $
    && $  \;\; (-1)^l,$ with & $ \;\; (-1)^l,$ with\\
&&&& $~l = 0, 2, \cdots, 2k (= n\!-\!1)$ & $~l = 0, 2, \cdots, 2k\!-\!2 (= n\!-\!2)$ \\
(\# allowed) &&$\;\; (1)$&& $  \;\;\; (k+1 = \frac{n+1}{2})$ &$ \;\; (k = \frac{n}{2})$ \\
\hline
%&&&&&\\
%\hline
$\overline{U} [\overline{P}^\alpha, \Delta^{\mu_1}] U$ & $ + $ &
$ \;\; - $
    && $  \;\; (-1)^l,$ with & $ \;\; (-1)^l,$ with\\
&&&& $~l = 0, 2, \cdots, 2k (= n\!-\!1)$ & $~l = 0, 2, \cdots, 2k\!-\!2 (= n\!-\!2)$ \\
(\# allowed) &&$\;\; (1)$&& $  \;\;\; (k+1 = \frac{n+1}{2})$ &$ \;\; (k = \frac{n}{2})$ \\
\hline \hline
%&&&&&\\
(Total \#) && $3$ && $\; 4k\!+\!3 = 2n\!+\!1$ & $ \; 4k = 2n$ \\
%&&&&&\\
\hline\hline
\end{tabular}
\end{center}
\end{table}

It is obvious that the two enumerations are consistent with each other, that is,
time reversal invariance does not further limit the number of independent form
factors. Therefore, the explicit decomposition of the non-forward matrix elements
of quark helicity-flip operators into independent general nucleon form factors,
Lorentz structures, and kinematical factors is \cite{ref:zc-count}
\beqar{eqn:Qexpansion}
    \bra{P'} {\cal O}^{\alpha \mu_1 \mu_2 \cdots \mu_n} \ket{P}
        &=& \overline{U}(P')\;\sigma^{\alpha\mu_1} \; U(P)
                \; \sum_{i=0}^{[\frac{n+1}{2}]}
            A_{n,2i\!-\!1} \; \Delta^{\mu_2}\Delta^{\mu_3} \cdots \Delta^{\mu_{2i\!-\!1}}
                \overline{P}^{\mu_{2i}} \cdots \overline{P}^{\mu_n} \non \\
        && + \; \overline{U}(P') \; [\gamma^\alpha, \overline{P}^{\mu_1}] \; U(P)
                \; \sum_{i=0}^{[\frac{n}{2}]}
            B_{n,2i} \; \Delta^{\mu_2}\Delta^{\mu_3} \cdots \Delta^{\mu_{2i}}
                \overline{P}^{\mu_{2i\!+\!1}} \cdots \overline{P}^{\mu_n} \non \\
        && + \; \overline{U}(P') \; [\gamma^\alpha, \Delta^{\mu_1}] \; U(P)
                \; \sum_{i=0}^{[\frac{n+1}{2}]}
            i C_{n,2i\!-\!1} \; \Delta^{\mu_2}\Delta^{\mu_3} \cdots \Delta^{\mu_{2i\!-\!1}}
                \overline{P}^{\mu_{2i}} \cdots \overline{P}^{\mu_n} \non \\
        && + \; \overline{U}(P') \; [\overline{P}^\alpha, \Delta^{\mu_1}] \; U(P)
                \; \sum_{i=0}^{[\frac{n+1}{2}]}
            i D_{n,2i\!-\!1} \; \Delta^{\mu_2}\Delta^{\mu_3} \cdots \Delta^{\mu_{2i\!-\!1}}
                \overline{P}^{\mu_{2i}} \cdots \overline{P}^{\mu_n} \;. \non \\
\eeqar

\section{Summary and Outlook}

The generalized parton distributions generalize both the Feynmen PDFs and form factors
and offer the possibility of explore at the parton level the three-dimensional structure,
and unravel the spin structure, of hadrons.

By using CPT invariance and crossing symmetry we obtained the enumeration of
independent general nucleon form factors of twist-two helicity-flip quark and
gluon operators. The numbers of independent form factors are: for spin-$J$
quark operator (rank $r=J$), $2J-1$ (odd $J$) and $2J-2$ (even $J$),
while for spin-$J$ gluon operator (rank $r=J+2$), $2J-2$ (odd $J$) and
$2J-1$ (even $J$). The form factor decomposition of the quark operators
is also explicitly written out, in terms of the independent form factors,
kinematical factors and Lorentz structures. Immediate further work would be
the explicit decomposition of the gluon helicity-flip operators.

\vspace*{1cm}

\acknowledgements

The author would like to thank the Overseas Chinese Physicist Association (OCPA)
and Haiyan Gao for invitation to the OCPA04 conference. The author is very
much benefited from collaboration and discussions with Xiangdong Ji. The author
would also like to thank Laura Wang, without whose encouragement and support
this work would not have been possible.

%\bibliography{protonref}

\end{document}